\documentclass{njcarticle}

\usepackage[all]{xy}
\usepackage{graphicx}

\usepackage{amssymb}
\usepackage{amsmath}
\usepackage{proof}

\newcommand{\ignore}[1]{}

\newcommand{\reffig}[1]{Figure~\ref{#1}}

\newcommand{\refthm}[1]{Theorem~\ref{#1}}

\newcommand{\refdefn}[1]{Defn.\ \ref{#1}}

\newcommand{\kernel}[1]{\mathrm{ker}(#1)}
\newcommand{\esc}{\widetilde{~~}}

\newcommand{\coding}[1]{\ulcorner #1 \urcorner}
\newcommand{\partialrightarrow}{\rightharpoonup}

\newcommand{\rewrite}{\rightarrow}
\newcommand{\rewrites}{\overset{*}{\rewrite}}

\newcommand{\unsafe}{\mathsf{unsafe}}

\DeclareMathSymbol{\N}{\mathbin}{AMSb}{"4E}

\title{Language embeddings that preserve staging and safety}

\author{Todd L. Veldhuizen, Chalmers University of Technology}

\begin{document}

\maketitle

\ignore{***
\begin{figure}
\hrule

\vspace{0.1in}

TO DO
\begin{enumerate}
\item Can I show that if a language is stage universal, we can simulate
another language with constant-time interpretative overhead, i.e., if
an algorithm runs in time $O(f(n))$ in language $A$, then it runs in
time $O(f(n))$ in language $B$?  (Probably not, e.g., log overhead from
representing memory.)
\item How is succinctness affected by embedding.
\item Any multi-level language can be effectively embedded in a
stage-universal language, without loss in succinctness.  (What is
the use of multi-level languages?)
\item Languages with congruent vs. noncongruent division.
\item Daniel's question about e.g., concurrency primitives and the
ability to simulate them.
\end{enumerate}
\hrule
\end{figure}
***}

\begin{abstract}
We study embeddings of programming languages into one another
that preserve what reductions take place at compile-time, i.e., staging.  
A certain condition --- what we call a
`Turing complete kernel' --- is sufficient for a language to be
\emph{stage-universal} in the sense that any language may be
embedded in it while preserving staging.  A similar line of
reasoning yields the notion of safety-preserving embeddings,
and a useful characterization of \emph{safety-universality}.
Languages universal with respect to staging and safety are
good candidates for realizing domain-specific
embedded languages (DSELs) and `active libraries' that provide
domain-specific optimizations and safety checks.
\end{abstract}

\section{Introduction}

Embeddings of programming languages into one another are useful in 
studying their relative power and, sometimes, finding languages that are 
\emph{universal} in some sense.  
Examples include Turing-reducibility for studying computability,
poly-time reductions for subrecursive languages \cite{Royer:1994},
and `structure-preserving' embeddings for expressiveness
\cite{deBoer:IC:1994,Felleisen:SCP:1991,Mitchell:SCP:1993,Matsushita:PhD:1998}.


To further a search for languages suited to realizing
domain-specific embedded languages (DSELS) 
\cite{Sandewall:CSUR:1978,Emanuelson:Lisp:1980,Hudak:CSUR:1996}
and ``active libraries,'' \cite{Czarnecki:GP:2000}
we propose stage-preserving embeddings as a tool to study
languages in which some evaluation or simplification is guaranteed to
take place at compile-time.  Such guarantees can be wielded to
realize domain-specific optimizations and safety checks.
The principal result shown here is that if a language has
what we call a `Turing-complete kernel,' it is universal in
the sense that any language may be embedded into it while
preserving staging and safety properties.  

\subsection{Some background on computability}

Throughout this paper we shall rely on some basic notions from
computability theory.  We say a set of natural numbers $S \subseteq \N$
is \emph{decidable} or equivalently $\Delta^0_1$ when
there exists a Turing machine that given as input any $x \in \N$ can decide
whether $x \in S$.  A set $S \subseteq \N$ is 
\emph{computably enumerable} or $\Sigma^0_1$ when
there exists a Turing machine that given input $x \in \N$ will
halt exactly when $x \in S$. (We follow the recommendation of
Soare \cite{Soare:BSL:1996} that the traditional term 
\emph{recursively enumerable}
be retired in favour of the more descriptive term
\emph{computably enumerable}.)
These notions extend
easily to sets of strings and terms by
employing an appropriate coding of objects by natural numbers.
For example, strings over a finite alphabet $\Lambda$ can be 
encoded by treating a string $x \in \Lambda^{*}$ as a base-$| \Lambda |$ natural
number; we may then speak of a set of strings over $\Lambda$
as computably enumerable or decidable.  
A function implemented by a computer is appropriately modelled
by a partial function, since the computation may fail to
terminate for some values of the domain.  A partial function 
$f : \N \partialrightarrow \N$ is
computably enumerable or $\Sigma^0_1$ when it is computable by
a Turing machine; in this case we say $f$ is a partial computable function.

\section{Stage-preserving embeddings}

The formalization of programming languages and compilers
is susceptible to fussiness, and to keep this
at bay I propose to be precise where it matters
and vague where it does not.
Let us adopt a grossly simplified view, typical
of computability, in which a programming language is merely a 
set of programs represented by binary strings.
One way to achieve this perspective is to view the program text
(a sequence of characters) as a single, large binary string.
We shall suppose
the programming languages of interest are all being compiled to
one implementation language $L_M$ --- the subscript $M$ suggesting 
a \emph{machine} language.  
To speak of
translations being semantics-preserving, we require that
$L_M$ comes paired with an equivalence $\sim$ on machine language
programs capturing some desired notion of program equivalence ---
the precise meaning of $\sim$ does not matter for our purposes.
For two programs $p,p' \in L_M$, we write $p \sim p'$ to
mean they \emph{do the same thing}.

We define programming languages in terms of their compilation
to $L_M$.

\begin{definition}
A programming language is a pair $(L_A, \phi_A)$ with
$L_A$ a decidable set of binary strings representing valid
programs, and $\phi_A : L_A \rightarrow L_M$
a compilation map required to be computably enumerable.
\end{definition}

Some languages have compilers that do not necessarily
terminate ---
C++ and MetaML are examples \cite{Manthey:BEATCS:2003,Taha:TCS:2000}.
For this reason compilers are appropriately modelled by computably
enumerable partial functions, rather than total functions.
To keep the notational convenience of total functions we employ the usual 
device of introducing a special element $\bot \in L_M$ to indicate a 
nonterminating compilation, and require that $\bot$ is in a singleton
equivalence class under $\sim$, i.e., $p \sim \bot$ if and only if $p = \bot$.

\ignore{***
\begin{figure}
\begin{equation*}
\begin{array}{ll}
L_M & \mbox{Implementation (machine) language} \\
L_A & \mbox{A language of interest} \\
L_u & \mbox{Some supposedly \emph{universal} language} \\
\sim & \mbox{Program equivalence over } L_M \\
\phi_A : L_A \rightarrow L_M & \mbox{Compiler from } L_A \mbox{ to } L_M \\
\end{array}
\end{equation*}
\caption{Notations used in the formalization.}
\end{figure}
***}

\begin{definition}
\label{defn:semantics-preserving}
A language embedding $e: L_A \rightarrow L_B$ is an injective
and computable function that is semantics-preserving, i.e.,
$\phi_A(p) \sim \phi_B(e p)$ for all $p \in L_A$.
\end{definition} 

\noindent
The typical scenario we shall consider is illustrated by this diagram:

\begin{equation}
\xymatrix @=0.5cm {
L_A \ar[rdd]_{\phi_A} \ar[rr]^{e} & & L_u \ar[ddl]^{\phi_u} \\ 
\\
    & L_M \\
}
\end{equation}
\noindent
We have two source languages $L_A$ and $L_u$, compilers
$\phi_A$ and $\phi_u$ for them, and we consider an
embedding $e : L_A \rightarrow L_u$.  We ask
when embeddings that preserve properties of interest
(semantics, staging, safety) exist.  The scenario of special
interest is when $L_u$ is some language purporting
to be `universal.'

\index{staging}
We use the notion of \emph{stages} to address compile-time computations
(cf. \cite{Jones:1993,Taha:TCS:2000}).
We are interested in embeddings that are stage-preserving:
if a computation occurs at compile time in language $L_A$,
then it occurs at compile time in language $L_u$.  This can be
conveniently addressed using the \emph{kernel} of the compiler.
\index{kernel}
Recall that the kernel of a map $\phi$ is:
\begin{equation}
\kernel{\phi} = \{ (p_1,p_2) ~|~ \phi(p_1)=\phi(p_2)\}
\end{equation}
\begin{figure}
\begin{align*}
\xymatrix{
L_A \text{ programs (source)} & ~~~ & L_M \text{ programs (target)} \\
x + 2 \ar[rrd]^\phi \\
x + \esc{(1+1)} \ar[rr]^\phi        &  & \text{IADD}~x~2\\
x + \esc{(1+(2-1))} \ar[rru]^\phi \\
y + 2  \ar[rr]^\phi                        & & \text{IADD}~y~2 \\
y + \esc{(4-2)} \ar[rru]^\phi
}
\end{align*}
\caption{\label{f:kernel}
Illustration of the kernel of a compiler $\phi$ for some terms in
a hypothetical staged language with the \emph{escape} operator $\esc{()}$.  
Expressions enclosed by $\esc{()}$ are evaluated at compile time.  The kernel 
gives equivalence classes of source programs that map to the same compiled 
program, in this case $\kernel{\phi}$ yields the equivalence classes
$\{ \{ x + 2 , x + \esc{(1+1)}, x + \esc{(1+(2-1))} \},
\{ y + 2, y + \esc{(4-2)} \} \}$.
}
\end{figure}

\noindent
The kernel of a compiler is a simple but versatile notion.
The kernel is an equivalence relation; every program in a kernel-equivalence
class compiles to the same target program.
Kernels capture staging ---
from the kernel one can deduce
what compile-time reductions take place.
For instance, a language whose compile-time evaluations are defined 
by a rewrite relation
$\rewrite$ must satisfy $\rewrites \, \subseteq \kernel{\phi}$,
where $\phi$ is its compiler (\reffig{f:kernel} shows an
example of some MetaML-like terms).
A useful analogy may be drawn to linear algebra, where
the kernel of a linear transformation yields its 
\emph{nullspace}.  When a vector is transformed,
every component lying in the nullspace is zeroed.
Analogously, any code \emph{lying in the kernel}
of the compiler `disappears' at compile-time.
Thus we can view the kernel as a \emph{staging specification}
and use it to formalize the notion of a
stage-preserving embedding.\footnote{The 
kernel is related to, but different
from, binding-time specifications (cf. \cite{Jones:CSUR:1996,Jones:1993}):
the kernel indicates which programs will compile to the same target program,
whereas binding-times indicate which terms are replaceable by constants.
These two ideas coincide in some situations, e.g., when programs
are terms, the compilation map is compositional, and only partial evaluation
is taking place.}

\label{s:stage-preserving}
\index{stage-preserving embedding}
\index{embedding!stage-preserving}
\begin{definition}
\label{defn:stage-preserving}
An embedding $e: L_A \rightarrow L_u$ is stage-preserving
when it satisfies
$(p_1,p_2) \in \kernel{\phi_A} ~\Rightarrow~ (ep_1,ep_2) \in \kernel{\phi_u}$.
\end{definition}

\begin{figure}
\begin{centering}
\includegraphics[scale=0.6]{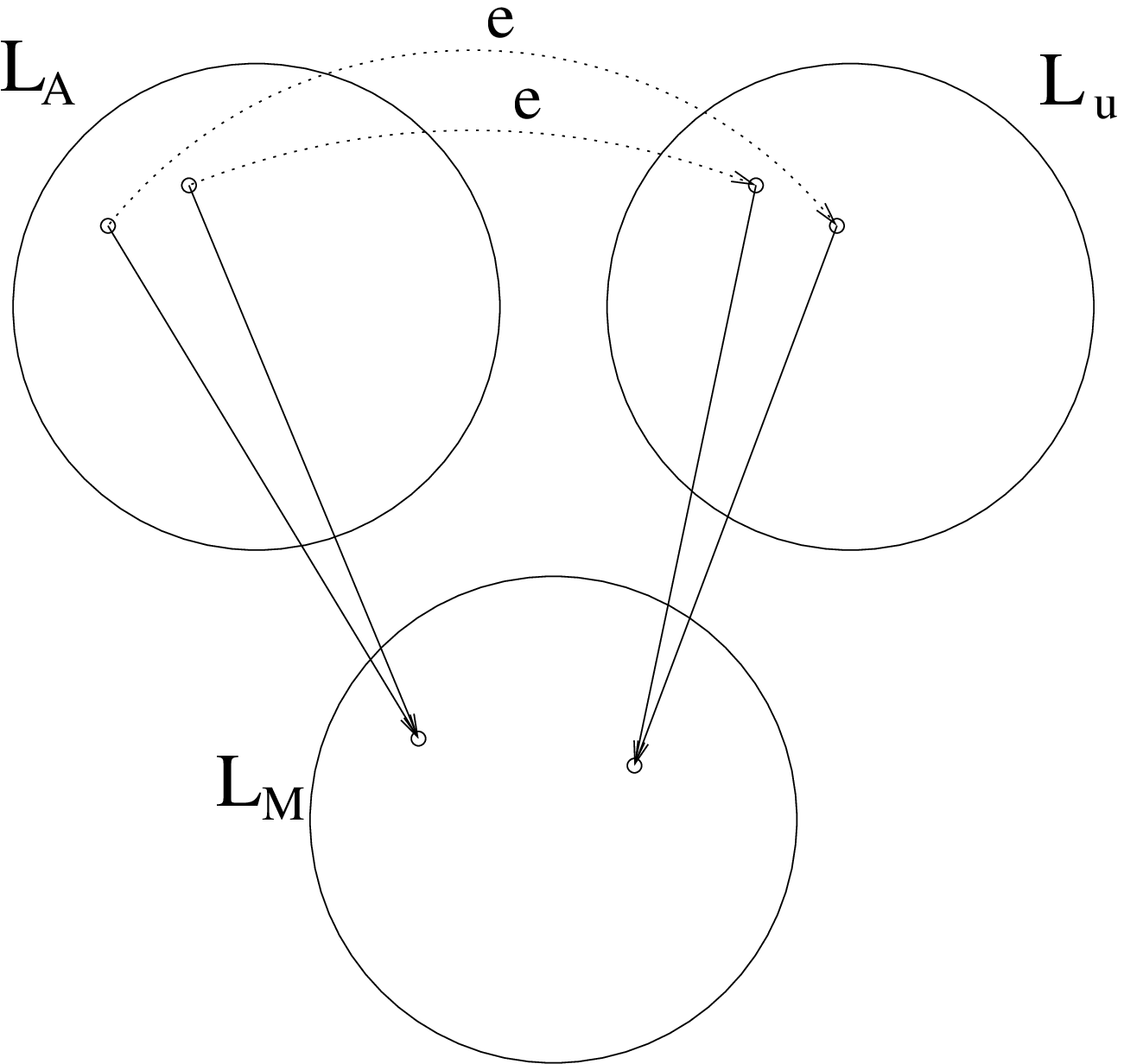}

\end{centering}
\caption{\label{f:stagepres}Illustration of stage-preserving embedding.
If two programs in $L_A$ compile to the same program in $L_M$, then 
after embedding in $L_u$ they must still compile to the same program.
Note, though, that it is not required that $\phi_A(p) = \phi_u(e p)$,
i.e., we do not expect to get the same target program going either
route, though this would be agreeable should it happen.
}
\end{figure}

\noindent
\label{s:kernel-staging-power}
\reffig{f:stagepres} illustrates.
The kernel of a compiler gives us a measure of its staging
power, that is, its ability to reduce computations at compile time.
\refdefn{defn:stage-preserving} effectively says: to increase
the staging power of a language, make its kernel larger.
But at what point is a kernel ``big enough'' that we can
embed any language into it and preserve staging?
To answer this, let us order languages, writing
$L_A \leq_S L_B$ to mean there exists a stage-preserving
embedding $e: L_A \rightarrow L_B$.  The relation $\leq_S$ is a preorder,
i.e., reflexive and transitive, but not necessarily anti-symmetric.
Given languages $L_A, L_B, L_C, L_D, \cdots$
we might have the following diagram of $\leq_S$,
with arrows indicating the existence of stage-preserving embeddings:
\begin{align*}
\xymatrix { 
  & \vdots \\
  & L_D \ar[u] \\
L_B \ar@/^1pc/[rr] \ar[ur] & & L_C \ar@/^1pc/[ll] \ar[ul] \\
  & L_A \ar[ul] \ar[ur]
}
\end{align*}
\noindent 
The obvious question is whether there might exist languages maximal in the
order $\leq_S$; we call such languages \emph{stage-universal}.

\begin{definition}
A programming language is stage-universal when there is a stage-preserving embedding
of any other programming language into it.
\end{definition}

\noindent
The term stage-complete would do equally well.  Now let us 
show that such languages exist and have a useful characterization.
We shall construct such a language and
refer to it as $L_u$, the subscript here indicating \emph{universal}.
The universal language $L_u$ is required to provide some standard
features of programming languages:
\begin{enumerate}
\item We assume there is
an effective coding $\coding{\cdot}$ of the languages
$L_A,L_M$ in $L_u$; this means we can
represent a program in $L_A$ by some term or computation
in the language $L_u$, and thereby examine and manipulate it.
If $p \in L_A$ is a program then $\coding{p}$ may be thought of
as a representation of $p$ by its parse tree, as a 
string of characters, or (more traditionally) a very large natural number;
the particulars do not matter so long as the encoding is unique
and computable.
\item We shall want to manipulate representations of programs in
$L_u$, so we assume $L_u$ permits the construction of functions
over codes (e.g., functions that manipulate parse trees), and 
write $F(c)$ to mean the application of such a function $F$
to a code $c$.  
It is useful to distinguish between functions implemented in $L_u$, e.g.,
purely functional manipulations of coded programs, and \emph{programs} 
such as interpreters
that take such codes and produce behaviour.  For a 
program $P$ taking as argument some code $x$, we write $P[x]$.  
\item We assume $L_u$ has function composition:
\begin{itemize}
\item If there are $L_u$-functions
$F$ and $G$, then there is an $L_u$-function $F \circ G$.
\item If there is a program $P[\cdot]$ and an $L_u$-function $F(\cdot)$,
then the construction $P[F(\cdot)]$ is meaningful:
there is some program $P_F[\cdot]$ such that $P_F[y] \sim P[x]$ when $x=F(y)$.
\end{itemize}
\end{enumerate}


\noindent
Much of what follows relies on the ability to interpret $L_M$ programs
in $L_u$.

\begin{definition}
An \emph{interpreter} for the machine language $L_M$ in
the language $L_u$ is a program $I_M[\cdot]$ such
that for every machine-language program $p_m \in L_M$, 
the interpreted version of $p_m$ is equivalent to $p_m$:
\begin{equation}
\phi_u(I_M[\coding{p_m}]) \sim p_m
\end{equation}
\end{definition}

\noindent
That is, if we take some machine-language program $p_m$ and
`code' it as (for example) a syntax tree $\coding{p_m}$
and give it to the interpreter $I_M$, then $I_M$ running $\coding{p_m}$
behaves the same way as the program $p_m$.  
The existence of such an interpreter ensures that the language $L_u$ does not 
lose basic capabilities of the language $L_M$, such as the
ability to interact with the operating system and so forth.
This is of concern when dealing with interactive
systems (a.k.a. processes, reactive systems, etc.)
rather than purely functional programs.  
The existence of such an interpreter
guarantees that $\phi_u$ is \emph{onto} the equivalence classes $L_u / \sim$
giving the possible behaviours
of $L_M$ programs.  
That is, for every machine-language program 
$p_m \in L_M$, there is a
program $p_u \in L_u$ such that $p_u$ is indistinguishable in behaviour
from $p_m$, i.e., $\phi_u(p_u) \sim p_m$.

What we need next is some vocabulary to discuss compile-time
computations in the language $L_u$. 
We work from the assumption
stated earlier that $L_u$ has a mechanism for defining functions.
\begin{definition}
A partial function $f$ is `\emph{realizable in the kernel}' of $\phi_u$ if
there exists an $L_u$ function $F$
such that for any program $P$ taking as argument a code,
and for any $x,y$ such that $y=f(x)$:
\begin{equation}
\phi_u(P[F(\coding{x})]) = \phi_u(P[\coding{y}])
\end{equation}
\noindent
Or, equivalently, $(P[F(\coding{x})],P[\coding{y}]) \in \kernel{\phi_u}$.
\end{definition}

\noindent
This means, more or less, that the partial function $F$ is evaluated at
compile time.

We now give a sufficient condition for stage-universality, inspired by
ideas from partial evaluation, in particular
Jones-optimality \cite{Jones:1993} and the Futamura
projections \cite{Futamura:SCC:1971}.
The proof is boilerplate computability theory and partial
evaluation.  We rely heavily on the assumption (stated earlier) that 
compilers are $\Sigma^0_1$ functions.

\begin{theorem}
\label{thm:complete-kernel}
If
\begin{enumerate}
\item[(i)] there is an interpreter $I_M[\cdot]$ for $L_M$ in $L_u$; and
\item[(ii)] any $\Sigma^0_1$ function $f$ is realizable in the kernel 
of $\phi_u$,
\end{enumerate}
then the language $L_u$ is stage-universal.
\end{theorem}
\begin{proof}
Pick a language and compiler $L_A$ and $\phi_A$.
Since $\phi_A$ is $\Sigma^0_1$, by (ii) there is a $L_u$-function 
$\Phi_A$ realizing
it such that if $p_m = \phi_a(p_a)$ then 
$\phi_u(P[\Phi_A(\coding{p_a})]) = \phi_u(P[\coding{p_m}])$
for any program $P$ taking a code-argument.

Consider the embedding $e : L_A \rightarrow L_u$ given by:
\begin{equation}
e(p_a) = I_M[\Phi_A(\coding{p_a})]
\end{equation}

\noindent
where $I_M[\cdot]$ is the $L_m$ interpreter whose existence is ensured
by (i).
Recall from \refdefn{defn:stage-preserving} that
$e$ is stage preserving when
$(p_1,p_2) \in \kernel{\phi_a} ~\Rightarrow (ep_1,ep_2) \in \kernel{\phi_u}$.
Choose $p_1,p_2$ such that $(p_1,p_2) \in \kernel{\phi_a}$.
Then there is a $p_m$
such that $\phi_a(p_1) = \phi_a(p_2) = p_m$, and from
the choice of $\Phi_A$,
\begin{equation}
\begin{array}{lcll}
\phi_u(I_M[\Phi_A(\coding{p_1})]) &=& \phi_u(I_M[\coding{p_m}]) & \mbox{and} \\
\phi_u(I_M[\Phi_A(\coding{p_2})]) &=& \phi_u(I_M[\coding{p_m}]) 
\end{array}
\end{equation}
Therefore $\phi_u(e p_1) = \phi_u(e p_2)$, or
$(ep_1,ep_2) \in \kernel{\phi_u}$, and
the embedding $e$ is stage-preserving.
Since such an embedding exists for any language $L_A$,
the language $L_u$ is stage-universal.
\end{proof}

We shall be sloppy henceforth and refer to a ``Turing-complete kernel''
to mean the properties listed in \refthm{thm:complete-kernel}.

The construction in the proof above is not of immediate practical
use; there is no guarantee that an interpreted program
$\phi_u(I_M[\Phi_A(\coding{p})])$ 
will run anywhere near as fast as $\phi_A(p)$
(cf. Jones-optimality \cite{Jones:1993}).
It does, however, give sufficient conditions for languages to
be stage-universal:
\begin{quote}
A language with a Turing-complete kernel can, in principle, subsume any
staged language.
\end{quote}

\noindent
This suggests we look to such languages to realize DSELs and
`active libraries.'
The construction above would be
useful if $\phi_u$ found programs that were `optimal.'  That is,
if the compiler $\phi_u$ were to find fastest, smallest, etc. programs, then
the construction $\phi_u(I_M[\Phi_A(\coding{p})])$ \emph{would}
be practical.  Finding optimal programs is undecidable, so
this goal is not reachable.
However, if we
find programs that are near to optimal, then approaches nearing
the construction of \refthm{thm:complete-kernel} might be practical.  
In \cite{Veldhuizen:2004} one possible method for realizing such
compilers is described, using ``Guaranteed Optimization,'' a new 
compiler design technique.

\section{Safety-preserving embeddings}

\label{s:safetyembed}
Let us now turn to the question of when there exist language
embeddings that preserve judgments about safety properties.  

Since useful safety properties are often undecidable, compilers approximate
the set of safe programs in a conservative way.  
For example, many compilers incorporate a static typing phase that
determines whether programs are well-typed in some formalism; programs that fail
typing are rejected since they might be unsafe.
When embedding one language into another, it is important that the set
of programs judged to be safe is preserved.  In particular we must
avoid the possibility that a language embedding might allow us to run
programs that fail safety checks in the source language.

In the real world, compilers react to programs they judge
unsafe by producing no output program and issuing a variety of diagnostic messages.
For ease of modelling, let us suppose compilers have one designated
output $\unsafe \in L_M$ signifying a program that fails safety checks.
The intent is that
a compiler $\phi$ judges a program $p$ to be unsafe exactly when
$\phi(p)=\unsafe$.  
There is then an obvious sense in which
an embedding can be safety-preserving.

\begin{definition}
\label{defn:safety-preserving}
A \emph{safety}-preserving embedding $e: L_A \rightarrow L_u$ is a
semantics-preserving embedding that preserves the set of programs
judged unsafe,
i.e., $\phi_u(e p)=\unsafe$ if and only if $\phi_A(p)=\unsafe$.
\end{definition}

We require that no programs in $L_M$ are equivalent to
$\mathsf{unsafe}$ \emph{except} $\mathsf{unsafe}$
itself, i.e., $\mathsf{unsafe}$ has a singleton equivalence class 
under $\sim$.
This means, incidentally, that semantics-preserving
(\refdefn{defn:semantics-preserving}) implies safety-preserving.

\ignore{***
Let us define the set of safe programs for a language $L_A$ to be
\begin{equation}
\mathsf{Safe}_A = \{ p \in L_A ~|~ \phi_A(p) \neq \unsafe \}
\end{equation}

Note this says nothing about a run-time safety property, merely
that such programs pass compile-time safety checks.
Such checks may be an implementation of a proof calculus
or something more informal, but crucially, if the safety check
is effective then $\mathsf{Safe}_A$ is a $\Sigma^0_1$ set.

***}

Following a similar line of reasoning as before,
we ask whether there are languages that
are safety-universal, in the sense that any language may be
embedded into it while preserving safety.
There are two approaches
we explore here.  The first is to note an obvious, but 
somewhat unenlightening, corollary of \refthm{thm:complete-kernel}:

\begin{corollary}
Any language meeting the criteria of \refthm{thm:complete-kernel}
is safety-universal.
\end{corollary}

This follows because stage-preserving embeddings are
semantics-preserv\-ing, and from the way we defined the
special compiler output $\unsafe$, any 
stage-preserving transformation is safety-preserving
(\refdefn{defn:safety-preserving}).  Therefore any
stage-universal language is also safety-universal.

For a more informative construction, let us consider compilers that employ 
a preliminary safety checking phase.  
We presume this safety checking phase implements a proof calculus $\vdash$ making
judgments of the form $\vdash \mathrm{safe}(p)$, indicating the
program $p$ is safe.  This is a general
framework that subsumes, for example, type systems;
we can augment a typical type inference system with an additional
rule of the form:
\begin{align*}
\infer{\vdash\mathrm{safe}(p)}{\vdash p : \tau}
\end{align*}
\noindent
This states that if a program $p$ can be given a type $\tau$, then
it is safe.  We limit ourselves to \emph{effective} proof calculi,
i.e., those whose deductions are computably enumerable,
and in particular to relations $\mathrm{safe}(p)$ that are decidable.
We will write $\not\vdash \mathrm{safe}(p)$ to mean
``$\mathrm{safe}(p)$ is not a valid deduction of $\vdash$.''

\begin{theorem}
\label{thm:safety-universal}
Let $L_A,\phi_A$ be a language and its compiler, and
$\vdash$ be an proof calculus with judgments
of the form $\vdash \mathrm{safe}(p)$ for some $p \in L_A$,
such that the set $\{ p ~|~ \vdash \mathrm{safe}(p)\}$ is decidable.
Let $L_u,\phi_u$ be a language and compiler meeting the criteria of
\refthm{thm:complete-kernel}.  Then there
is a stage-preserving embedding $e : L_A \rightarrow L_u$ such that
$\phi_u(e p) = \mathsf{unsafe}$ if and only if $\not\vdash \mathrm{safe}(p)$.
\end{theorem}

\begin{proof}
Consider the function $\phi_A': L_A \rightarrow L_M$
given by:
\begin{align*}
\phi_A'(p) = 
  \begin{cases}
  \phi_A(p) & \text{if}~\vdash \mathrm{safe}(p) \\
  \mathsf{unsafe} & \text{if}~\not\vdash \mathrm{safe}(p)
  \end{cases}
\end{align*}
\noindent
Since the set $\{ p ~|~ \vdash \mathrm{safe}(p)\}$ is decidable,
i.e. $\Delta^0_1$, and $\phi_A$ is $\Sigma^0_1$, the function $\phi_A'$
is $\Sigma^0_1$.  By the conditions of \refthm{thm:complete-kernel}, there
exists a u-function $\Phi'_A$ realizing $\phi_A'$ in the
kernel of $\phi_u$.  Consider the
embedding
\begin{align*}
e(p_a) &= I_M[\Phi_A'(\coding{p_a})]
\end{align*}
\noindent
Following the reasoning given in the proof of \refthm{thm:complete-kernel},
$e(p_a) = \unsafe$ if and only if $\not\vdash \mathrm{safe}(p)$,
and $e$ is a stage-preserving embedding.
\end{proof}
A key requirement, implicit in the above proof, is that the function
$\Phi_A'$ must be able to produce $\coding{\unsafe}$, i.e.,
the code of an unsafe program.  The intuition we can draw from
this is the following:

\begin{quote}
Any language with a Turing-complete kernel \emph{and} the ability
to construct at compile-time a condition signifying ``unsafe program''
is safety-universal.
\end{quote}

\section{Conclusions}

Variations on extensible and universal programming languages have
been explored for decades.  We have examined a new twist on this
theme, looking not just to languages that are \emph{Turing-complete}
(can perform any effective procedure)
or syntactically extensible (can provide a domain-specific syntax),
but to languages that are universal with respect to \emph{staging}
and \emph{safety}.  Such languages appear ideal for expressing
domain-specific safety checks and optimizations, suggesting a route
to realizing libraries and DSELs that are not only expressive, but also fast 
and safe.  

\bibliographystyle{njcarticle}

\end{document}